\documentclass[useAMS,usenatbib,usegraphicx]{mn2e}
\newcommand{\nhsym}{N_{\mbox{\scriptsize H}}}

\title[The X-ray absorber in Mrk 915: {\emph{Swift}} data]
{The structure of the X-ray absorber in Mrk 915 revealed by {\it Swift}}
\author[P. Severgnini et al.]
{P. Severgnini$^{1}$\thanks{E-mail:
paola.severgnini@brera.inaf.it},
L. Ballo$^{1}$, 
V. Braito$^{1,2}$, 
A. Caccianiga$^{1}$,
 S. Campana$^{1}$,
 \newauthor
R. Della Ceca$^{1}$, 
A. Moretti$^{1}$
and C. Vignali$^{3}$\\
$^{1}$INAF--Osservatorio Astronomico di Brera, via Brera 28, I-20121 Milano, Italy\\
$^{2}$Department of Physics, University of Maryland, Baltimore County, Baltimore, MD 21250, USA\\
$^{3}$Dipartimento di Fisica e Astronomia, Universit$\grave{a}$ degli Studi di Bologna, Viale Berti Pichat 6/2, I-40127 Bologna, Italy}
\begin{document}

\date{Accepted 2015 August 10. Received 2015 June 22, in original form 2015 April 2}

\pagerange{\pageref{firstpage}--\pageref{lastpage}} \pubyear{XXXX}

\maketitle

\label{firstpage}

\begin{abstract}
In this paper we present the results obtained with a monitoring programme (23 days
long) performed with  {\it Swift}-XRT on the local Seyfert galaxy  Mrk 915. The
light-curve analysis shows a significant count rate variation 
(about a factor of 2--3) on a time-scale of a few days, while
the X-ray colours  show  a change in the spectral curvature below 2 keV and the
presence of two main spectral states. 
From the spectral analysis we find that the observed variations
can be explained by the
change of the intrinsic nuclear power (about a factor of 1.5) coupled with a change
of the properties of an ionized absorber.
The quality of the data prevents us from firmly establishing if the
spectral variation  is due to a change in the ionization state and/or in the 
covering factor of the absorbing medium. The latter scenario would imply a
clumpy  structure of the ionized medium. 
By combining the information provided
by the light curve and the spectral  analyses, we can derive some 
constraints on the location of the absorber under the hypotheses of either  homogeneous
or clumpy medium.  In both cases, we find that the absorber should be located
inside the outer edge of an extended torus and, in particular,  under the clumpy
hypothesis, it should be located near, or just outside, to the broad emission
line region.
\end{abstract}

\begin{keywords}
galaxies: active -- galaxies: individual: Mrk 915 --X-rays: galaxies.
\end{keywords}

\section{Introduction}

There is now a general consensus that active galactic nuclei (AGN) are powered
by accretion of matter on to a super-massive ($>$10$^6$ M$_{\odot}$) black hole
(SMBH). According to the unified model of AGN \citep{Ant93}, an obscuring
optically thick medium composed by dust and gas arranged in a torus-like
geometry is present around the nuclear engine. However, the structure, size and
composition of this circumnuclear medium are still matter of debate and are
topics of several studies carried out at different wavelengths \citep{Bia12}. Important
constraints on the physical properties of the circumnuclear medium have been
recently provided by the study of the absorption variability, which is almost
ubiquitous in bright absorbed AGN \citep{Ris02}. X-ray absorbing column density
($\nhsym$) has been observed to vary by a factor of 10 or more over a few years.
The observed $\nhsym$ variability on this scale ruled out the simplest physical
configuration of a homogeneous absorber, giving upper limits on the distance of
the latter by the central SMBH. Subsequent X-ray observational campaigns
detected significant variability on a handful of AGN on very short time-scale
(day or week) giving the possibility to  investigate the X-ray properties of
the circumnuclear medium down to sub-parsec scale. The emerging picture is that
multiple neutral and ionized absorbers co-exist around  the central SMBH,
located at different distances from it. So far the best characterization of the
physical parameters of the X-ray absorbers has been possible for the AGN in NGC
1365, which has been monitored several times in the last few years with {\it Chandra},
{\it XMM-Newton} and {\it Suzaku}.  This source changes from Compton-thick
($\nhsym$$\geq$10$^{24}$ cm$^{-2}$) to Compton-thin ($\nhsym$$\sim$10$^{23}$
cm$^{-2}$) state on time-scales from weeks \citep{Ris05} to hours or days
\citep{Ris07,Ris09}.   One of the main results is that the X-ray absorber has
the physical properties typical of  the clouds responsible for the emission of
the broad lines in the optical/UV spectrum, i.e. the broad-line emission region
(BLR), located at a distance of hundreds of gravitational radii from the
central SMBH. 
Recently it has been proposed that the long-term variability of the X-ray absorber
in NGC~1365 could be due to the variation of the ionized X-ray wind, 
that is responding to the changes in the accretion rate \citep{Con14}.
High spectral resolution data show that this wind is located within the variable
soft X-ray absorber and is composed by two (or even three) zones with different
ionization levels. In particular, the lowest ionization zone of this wind could be
responsible for the absorption variability that is observed in this source below 2 keV \citep{Bra14}.

By exploiting the {\it Swift}  X-ray Telescope (hereafter XRT; \citealt{Bur05}) 
archive, we recently started a project aimed at    studying, on a larger
statistical basis, the physical properties of the X-ray absorbers in AGN. We
started from the 70 months  all-sky survey {\it Swift}-BAT catalogue \citep{Bau13}
in the 20--150 keV and the 66-month Palermo BAT
Catalogue\footnote{http://bat.ifc.inaf.it/bat\_catalog\_web/66m\_bat\_catalog.html}
(see also \citealt{Cus10}). We considered only those sources  with a secure optical AGN
counterpart and  with XRT observations  already available before the end of
2012. For each source, we adopted a daily binning of the XRT data and we
searched for statistically significant variations of the X-ray colours on
different time-scales (from months to years, see \citealt{Bal15}). 
Following this method, we selected five new candidates for variable
absorbers. One of the most
interesting is the local Seyfert galaxy Mrk 915.  The archival XRT data used to
select this source unveiled a dramatic count rate and `X-ray colour' variability
on a minimum time-scale of 1 month. The 0.3--10 keV
count rate increases by a factor of 3  and, at the same time, the spectral shape
becomes dramatically softer, showing a variability of a factor of 2.4 and 3 in
the  [2--10 keV]/[0.3--2 keV] and [4--10 keV]/[0.3--4 keV] count rate ratios,
respectively.  As discussed in \citet{Bal15}, this behaviour suggests
that a change in the properties of the X-ray obscuring circumnuclear medium (i.e.
column density, covering factor or ionization state)  probably occurred
between the archival XRT observations. 
However, the low statistical quality of
these data prevented us from performing a proper spectral analysis and to
discern among the possible causes of the observed variation.
With the goal of obtaining higher quality data on a shorter time-scale, we
performed a daily XRT monitoring on this source during a period of $\sim$3 weeks.
This paper aims at providing a first characterization of the putative variable
absorber.

The paper is structured in the following way: the main optical properties of Mrk~915 are reported in Section
2, while the analysis of the XRT data obtained during the daily monitoring are presented in Section 3 (light-curve 
analysis and spectral analysis, Sections 3.1 and 3.2, respectively). The main results obtained are
discussed in Section 4 and the conclusions are  summarized in Section 5.

Throughout the paper we assume a flat  $\Lambda$CDM cosmology with H$_0$=71 km s$^{-1}$ Mpc$^{-1}$, 
$\Omega$$_\Lambda$=0.7 and $\Omega$$_{\rm{M}}$=0.3.

\section{Mrk 915} 

\begin{table}
 \centering
  \caption{Mrk 915: XRT monitoring observation log. The observations are ordered on
  the basis of the observation start date.}
  \label{obs_log}
  \begin{tabular}{@{}lccc@{}}
  \hline
  \hline
Obs. ID & Obs.  start date& Net counts  &  Net exp. time\\
        &   & [0.3--10 keV]        &      [s] \\
 \hline
 00035169003 &  2014-09-10 & 638 & 7367 \\
 00035169004 &  2014-09-11 & 621 & 7060\\
 00035169005 &  2014-09-15 & 344 & 4453 \\
 00035169006 &  2014-09-17 & 1359 & 13130 \\
 00035169007 &  2014-09-18 & 437 & 3629\\
 00035169008 &  2014-09-21 & 2563 & 13190 \\
 00035169009 &  2014-09-23 & 693 & 4105\\
 00035169013 &  2014-09-27 & 160 & 1171 \\ 
 00035169014 &  2014-09-27 & 61 & 719\\ 
 00035169011 &  2014-09-29 & 935 & 10280\\
 00035169015 &  2014-10-02 & 993 & 5369\\
 00035169016 &  2014-10-02 & 1216 & 5314\\
\hline
\end{tabular}
\end{table}
\begin{figure*}
\centering
\includegraphics[width=14.5cm,height=5.5cm]{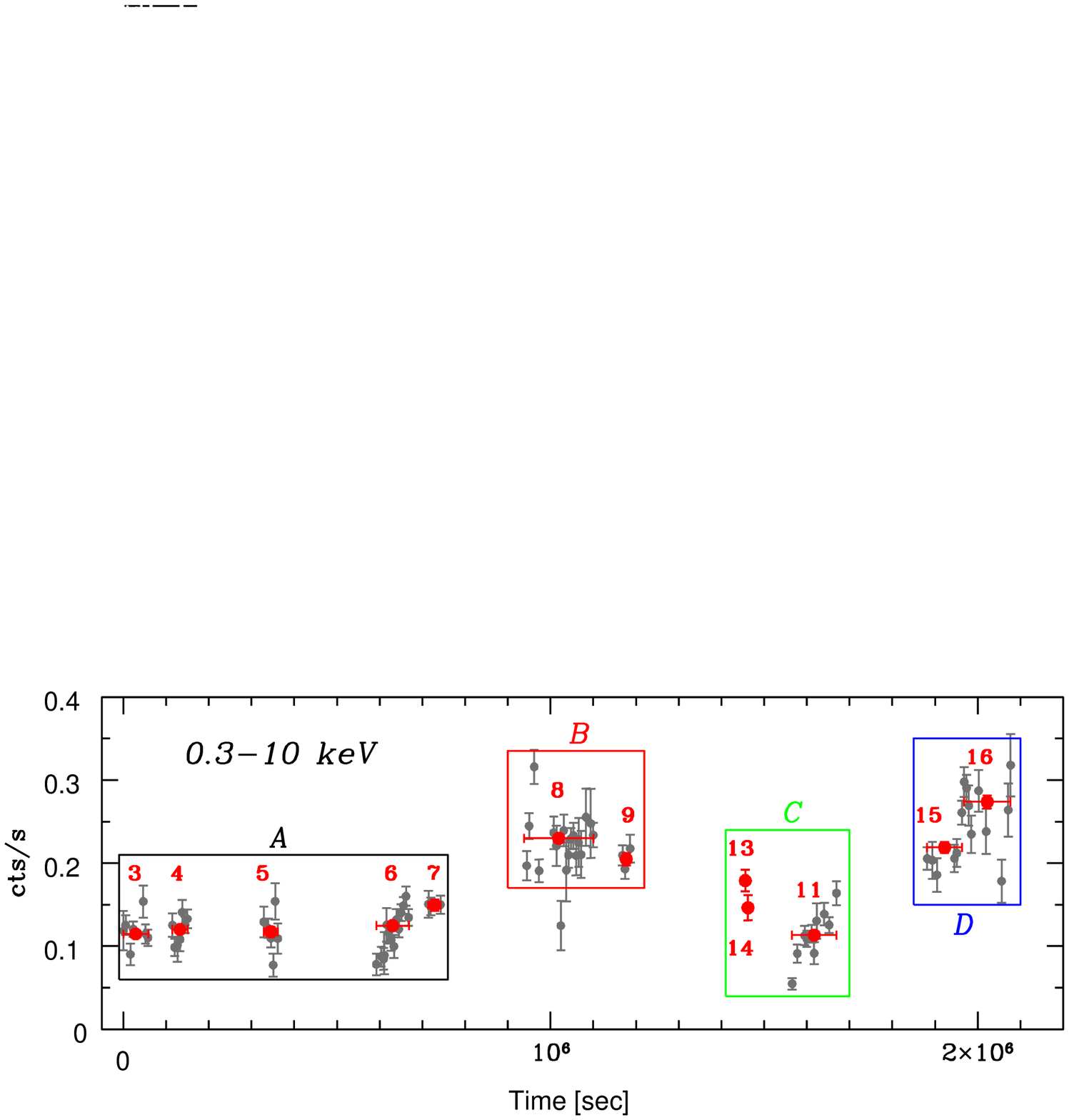}
\caption{XRT light curve in the total 0.3--10 keV band,  
obtained by binning the data per snapshot 
(small dots, grey in the electronic version) and  per observation (big dots, red in the electronic version). 
With respect to the observed net counts reported in Table~\ref{obs_log}, data points are corrected 
for technical issues (i.e. 
bad pixels/columns, field of view effects, pile-up and source counts landing outside the 
extraction region) following the receipts discussed by \citet{Eva07} and \citet{Eva09}.
Numbers mark the one/two last digital numbers of the relevant observation ID (see Table~\ref{obs_log}).
Error bars mark 1$\sigma$ uncertainties. The light curve is divided in time  intervals marked by different boxes 
and capital letters (see Section 3.2). We note that the only point in the  {\it B} time range with a
count rate lower than 0.2 has a very short exposure time ($\sim$180 s). Although
it is above the minimum exposure time considered here (i.e. 150 s), 
this point has a signal--to--noise ratio lower than 5.}
\label{fig1}
\end{figure*}

\begin{figure*}
\centering
\hskip -1cm\includegraphics[width=9cm]{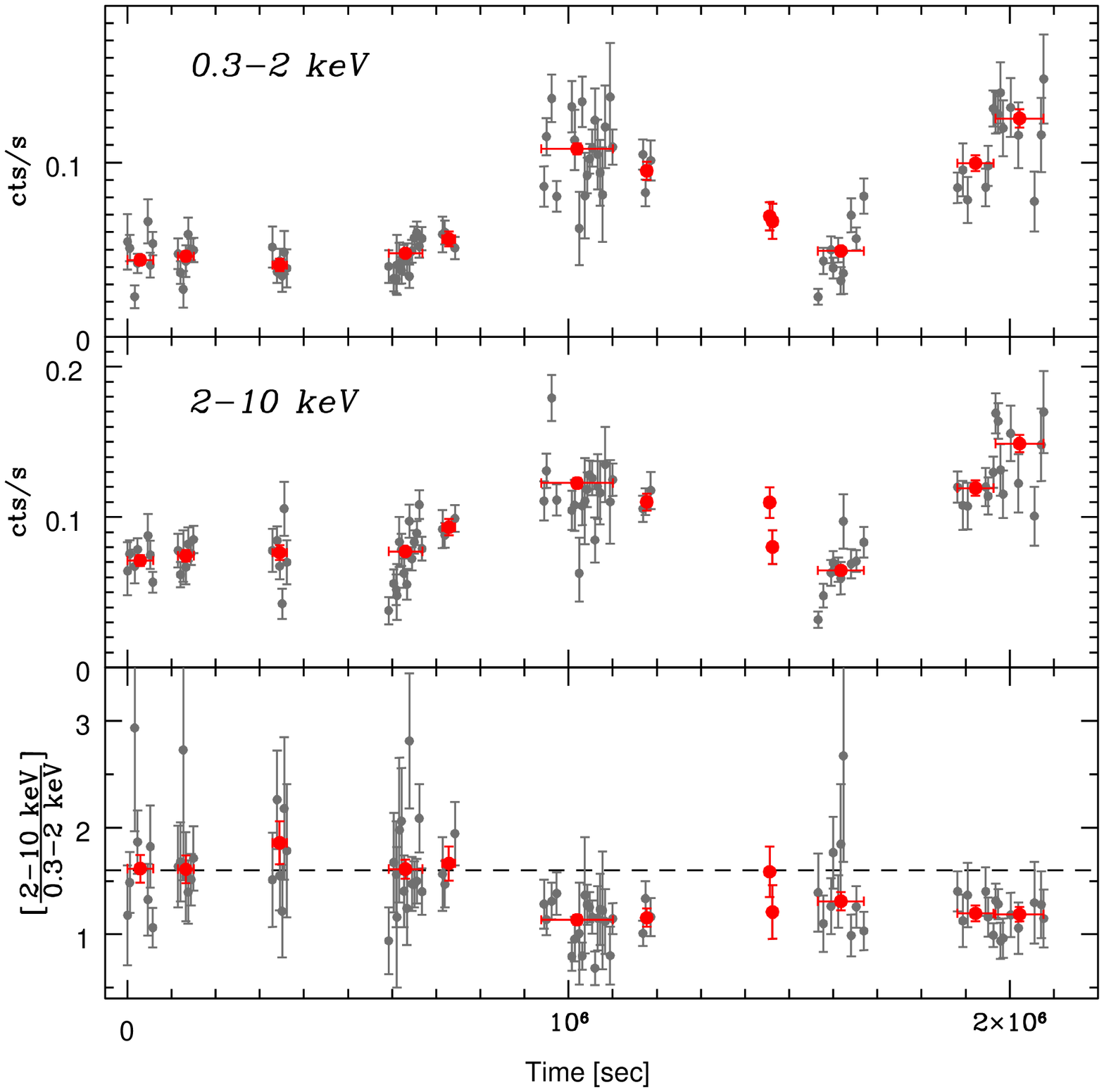}
\hskip 0.1cm\includegraphics[width=9cm]{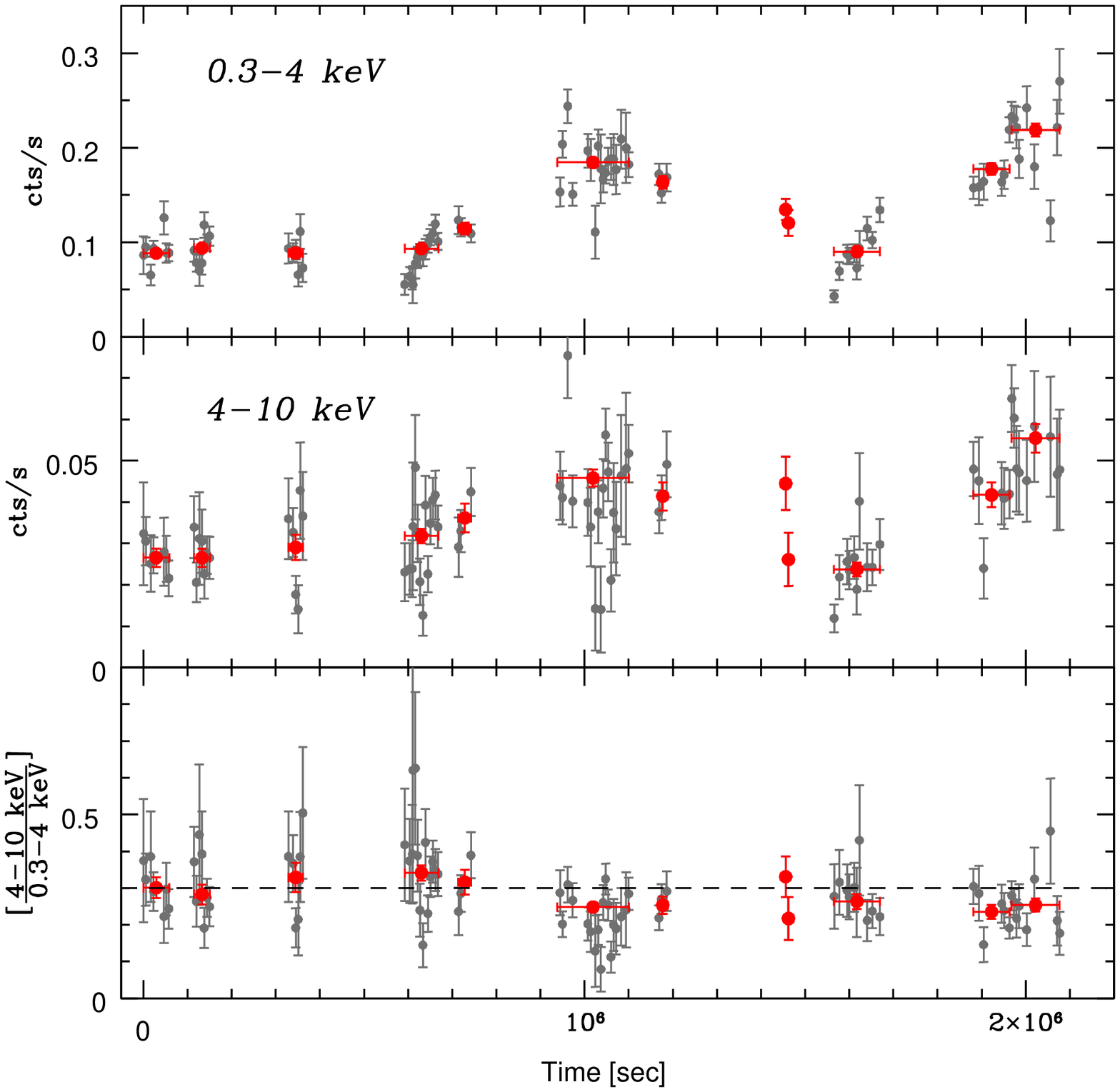}
\caption{
Left panels, from top to bottom: 0.3--2 keV and 2--10 keV count rates and [2--10 keV]/[0.3--2 keV] 
ratio. Right panels, from top to bottom: 0.3--4 keV and 4--10 keV count rates and 
[4--10 keV]/[0.3--4 keV] ratio. Dashed lines represent the ratios 
normalized to the first observation. 
Symbols are the same as those used in Fig.~\ref{fig1} and data points are corrected for the same technical
issues.}
\label{fig2}
\end{figure*}

Mrk 915 is a local ({\it z}=0.024) spiral galaxy (PA$\sim$166$^o$) with evident dust lane structures seen
crossing the central source  in its WFPC2 image \citep{Mal98} and  hosting an
SMBH of $M_{\rm BH}$=(0.6-1.8)$\times$10$^8$ M$_{\odot}$
\citep{Ben06}.  The value of the axis ratio (0.54 and 0.83 reported by \citealt{Kee96}
and \citealt{Mun07}, respectively) indicates that the disc of
the galaxy is inclined with respect to the line of sight by about
35$^{o}$-57$^{o}$.  Assuming both galaxy's disc and AGN's torus to be co--planar
and an  half-opening angle of the torus of about 60$^o$,
our line of sight intercepts
or grazes the low--density outer regions of the torus. 

From the optical point of view, Mrk~915 has been spectroscopically classified as an
intermediate Seyfert\footnote{A Seyfert galaxy  with broad emission  line components weaker than
those usually observed in Seyfert 1 galaxies \citep{Whi92}.} galaxy 
showing significant spectral variations.
In particular, the broad components of the H$\alpha$ and H$\beta$ emission lines 
are clearly 
detected in some observations, while they are completely absent in 
others. As a consequence, the source has been classified in different 
ways, from Seyfert 1.5 to Seyfert 1.9, depending on the presence/intensity of the 
broad components at the time of the observations \citep{Goo95,Ben06, Tri10}. 
The origin of this spectral variability is still unclear.
\citet{Goo95} tentatively suggested that the observed variation was due to 
a change in reddening ($\Delta$E$_{B-V}$$\geq$0.53 mag, derived by comparing
the H$\alpha$ fluxes in the two observations) produced by
dusty clouds passing close, but outside, the bulk of the BLR. 
Alternatively, the observed spectral variability could be due either to a 
change in the nuclear photoionizing continuum or to a combination of
continuum and reddening variation.
At the moment, it is not possible to
discriminate among the various scenarios \citep{Tri10}.

Independently of the origin of this spectral variability, a reddening of the
order of E$_{B-V}$$\sim$0.3-0.5 mag seems to be always present and it affects
both the kpc-distance regions, i.e. the  narrow-line emission regions (NLR),
and the innermost nucleus. This reddening is most probably 
attributed to the presence of dust lanes as seen crossing the central source
in its WFPC2 image \citep{Mun07}.

\section{Daily/weekly X-ray monitoring}

The data presented here are relative to a monitoring programme composed by 12 new
XRT observations awarded during fall 2014 ({\it Swift Cycle--10} and ToO
observations, P.I. Severgnini). These observations, covering $\sim$23 days, for a
total on source exposure time of $\sim$76 ks, were performed with XRT
in the standard PC-mode. The observation log is reported
in Table~\ref{obs_log}. The source appears  point--like in the XRT image and we do not find
any significant evidence of pile-up.

\subsection{Light curves} 

Each light curve, described below, has been created by using the  {\it Swift}-XRT
data products generator\footnote{http://www.swift.ac.uk/user\_objects/ }
developed by the UK {\it Swift} Science Data Centre and based on the Swift
software and \textsc{ftools} guide. The effects of the damage to the CCD and automatic
readout-mode switching are appropriately handled; details on the light-curve
code can be found in \citet{Eva07} and \citet{Eva09}.

Figs~\ref{fig1} and \ref{fig2} show the XRT light curves in five different energy bands
(0.3--10, 0.3--2, 0.3--4, 2--10 and 4--10 keV). We choose to bin
the data per spacecraft orbit (i.e snapshot, grey small dots in Figs~\ref{fig1} and \ref{fig2}).
We considered only those snapshots longer than 150 seconds, which guarantee  a
signal-to-noise ratio on each point equal or greater than 5. In each panel, we
plot also the light curves obtained by binning the data per observation (i.e.
obs. ID, big red dots in Figs~\ref{fig1} and \ref{fig2}).  Over the length of the
observing monitoring programme ($\sim$23 days), the light curves are not constant at
99.9 per cent confidence level ($\chi^2$ test).

During the first three observations, spanning $\sim$3.6$\times$10$^{5}$ s
(i.e. 4.2 days), the snapshot 0.3--10 keV count rates are randomly distributed
around $\sim$0.12 counts s$^{-1}$. Later, a clear increase in the count rate is visible,
reaching its maximum in about 3.5$\times$10$^{5}$ s (i.e. $\sim$4 days).  In
these four  days, the count rate increased  by a factor $\sim$2--3 when we consider
 the obs. ID time binning. About 2 days later, the count rate  decreased by the same amount
 and  then immediately increased again, reaching the highest intensities
observed during  this monitoring ($\sim$0.3 counts s$^{-1}$).

The increasing/decreasing factor is not the same in all bands: in the softer
bands (0.3--2 keV and 0.3--4 keV) the count rate variation is higher with respect
to that in the complementary harder bands (2--10 keV and 4--10 keV,
respectively). This is visible also inspecting the flux ratios ([2--10 keV/0.3--2 keV]
and  [4--10 keV/0.3--4 keV]) reported in the bottom panels of Fig.~\ref{fig2}. In these
panels, the dashed lines are normalized to the first observation. This behaviour
suggests that, besides the possible flux variation of the primary emission, 
there could be  another mechanism producing different intensity variations in
different energy bands. This  mechanism  should affect mainly the softer energy
ranges. 

\begin{figure}
\includegraphics[width=5cm, height=8.7cm, angle=-90]{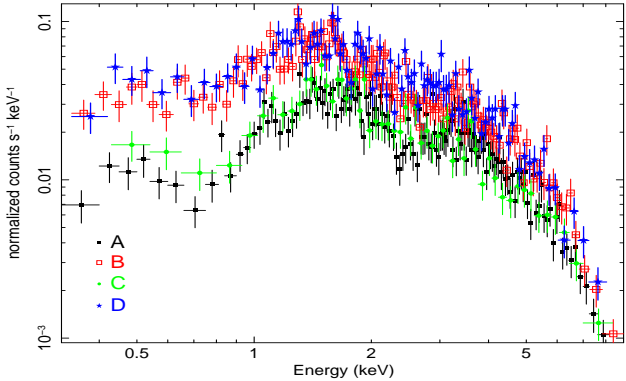}
\caption{XRT spectra relevant to the time binning defined in Fig.~\ref{fig1}. 
In the electronic version, the spectra are colour-coded as 
the boxes in Fig.~\ref{fig1}.}
\label{fig3}
\end{figure}

\subsection{X-ray spectral analysis}

\begin{table*}
 \centering
\begin{minipage}[h]{0.75\textwidth}
  \caption{Best-fitting values obtained by applying 
  to the low- and high-state spectra of Mrk 915 the following model:
  \textit{phabs}$\times$(\textit{zphabs}$\times$\textit{zxipcf}$\times$\textit{zpowerlw}) with $\Gamma$ fixed to 1.7 
  ($\chi$$^2/$d.o.f.=362.99/403).}
\label{model}
  \begin{tabular}{lccccccc}
  \hline
  \hline
State   &  $\nhsym$$^{a}_{\rm{neutral}}$  & $\nhsym$$^b_{\rm{ionized}}$ & CF$^c$  & log($\xi$)$^d$  &  $A^{e}_{\rm{p}}$ & $F_{\rm (2-10 keV)}$$^{f}$ & $L_{\rm (2-10
keV)}$$^{g}$\\
 \hline
 AC &  0.08$^{+0.03}_{-0.06}$ & 1.86$\pm{0.20}$ & 0.96$\pm{0.03}$           & 0.68$\pm{0.13}$        & 1.98$\pm{0.01}$ & 0.67 & 0.98\\ 
 BD &  0.08$^{h}$ & 2.09$\pm{0.35}$ & 0.86$^{+0.07}_{-0.05}$    & 1.14$^{+0.14}_{-0.24}$ & 3.06$\pm{0.01}$ & 1.07 & 1.52 \\
 \hline
 \hline
\end{tabular}
{\bf Notes:} The uncertainties, reported at the 90 per cent confidence level for one parameter of interest,
have been estimated by freezing the $\nhsym$$_{\rm{neutral}}$ to the best-fitting value.\\
$^a$ Column density of the neutral absorber associated with dust lane structures, 
in units of 10$^{22}$~cm$^{-2}$. This parameter is forced to be equal in both states.
$^b$ Column density of the ionized absorber, in units of 10$^{22}$~cm$^{-2}$.
$^c$ Covering factor of the neutral absorber.
$^d$ Logarithm value of the ionization parameter (see Section 3.2).
$^e$ Normalization of the power-law component, in units of 10$^{-3}$ photons~s$^{-1}$~cm$^{-2}$~keV$^{-1}$.
$^f$ Flux in the 2-10 keV energy range, corrected for the Galactic absorption, in units of 10$^{-11}$ erg cm$^{-2}$ s$^{-1}$.
$^g$ Intrinsic 2--10 keV luminosity, in units of 10$^{43}$ erg s$^{-1}$.
$^{h}$ This parameter was kept tied in the two states.\\
\end{minipage}
\end{table*}

The aim of this section is to investigate, from a spectral point of view, the
changes observed in the light curves of Mrk 915 during the X-ray monitoring.
To this end, following the total (0.3--10 keV) light curve, we divided the XRT
data in four main time bins on the basis of the average count rate value of each
observation.  They are shown in Fig.~\ref{fig1}, where we mark each
bin with boxes and capital letters. The A and C bins are composed by
observations with a net count rate lower than $\sim$0.2 counts s$^{-1}$, while B and D
correspond to observations with a net count rate higher than $\sim$0.2 counts s$^{-1}$.
For each bin, we extracted the spectrum by using circular regions centred on
the X-ray source position  with a radius $\sim$50 arcsec (i.e. $\sim$20
pixel, which corresponds to an  Encircled Energy Fraction of $\sim$90 per cent; \citealt{Mor05}). 
The background spectra have been extracted from source-free
circular regions close to the object with an area about 9 times larger.
Spectral reduction was performed using the standard
software (\textsc{headas} software, v6.15 and the most updated CALDB
version\footnote{http://heasarc.gsfc.nasa.gov/FTP/caldb})  and following the
procedures described in the instrument user
guide\footnote{http://heasarc.nasa.gov/docs/swift/analysis/documentation}.  
The spectra have been binned in order to have at least 20 counts per
energy channel (see Fig.~\ref{fig3}) and analysed using the \textsc{xspec} 12.8.1 package.

As evident from Fig.~\ref{fig3}, the
data have two by two (A with C and B with D) very similar spectral shape,
confirming the presence of two different spectral states of the source,  as
already highlighted by the light curve and X-ray colours. A visual inspection of
the figure is sufficient to notice that most of the spectral variation among these
two states is  below 2--3 keV. 
In order to improve the statistics, we combined the data sets with similar
spectral shape (A with C and B with D) by obtaining a low and a high--state
spectrum (hereafter called AC and BD, respectively). 

We also considered the {\it Swift}--BAT spectrum of Mrk
915 retrieved by the 70 month {\it Swift}-BAT
Catalogue\footnote{http://swift.gsfc.nasa.gov/results/bs70mon/} \citep{Bau13}. 
This spectrum provides the average spectral properties of the source above
$\sim$15 keV. Due to the significant variability of Mrk 915, we did not 
fit the XRT and BAT spectra simultaneously. We used the BAT spectrum 
to derive the average value of  the spectral slope
of the primary  AGN  emission. By fitting the BAT data with a simple power-law
component, we found a best-fitting value of $\Gamma$=1.7$\pm{0.2}$. 
In the spectral analysis of the XRT data (described in details below),
we first fitted simultaneously the two main states keeping $\Gamma$ tied.
Also in this case we found: $\Gamma = 1.7 \pm 0.2$. The same value and
uncertainties are obtained when $\Gamma$ is allowed to vary between the
states. Thus, considering the statistic of the present data, we decided to keep it
fixed to the best-fitting value while investigating the origin of the variability.

\begin{figure}
\centering
\hskip -0.7cm\includegraphics[width=5cm, height=9cm, angle=-90]{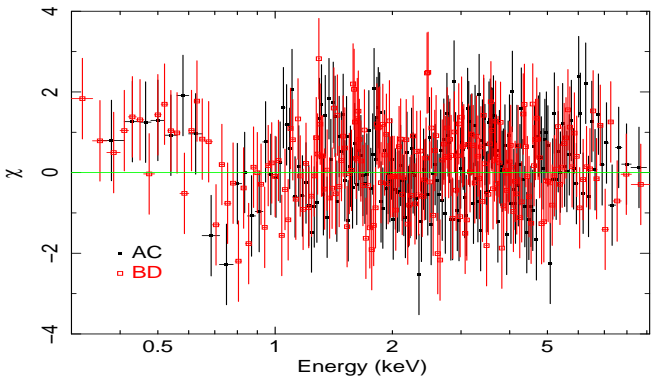}
\caption{Residuals, plotted in terms of sigmas with error bars of size one, obtained by
fitting the low- (AC) and high-state  (BD) spectra of Mrk~915 with 
the following model: {\it phabs}$\times$({\it zphabs}$\times${\it zpcfabs}$\times${\it zpowerlw}).}
\label{fig4}
\end{figure}

As a starting point, we assumed the presence of three absorbing components. The first one ({\it phabs}
component in {\textsc{xspec}})  refers to the Galactic hydrogen column density along the
line of sight \citep{Kal05}. The second one ({\it zphabs} component in {\textsc{xspec}})
accounts for the absorption due to the dust lane structures  at the redshift of the
source.  Indeed, as discussed in Section 2, the primary emission from the Mrk 915 nucleus
intercepts dust lane structures that produce an optical extinction of $E_{B-V}$=0.3--0.5 mag \citep{Ben06,Tri10}. The presence of these structures is
expected to affect also the X-ray emission with a column density of the order
of 10$^{21}$ cm$^{-2}$, assuming a Galactic dust-to-gas ratio of
E$_{B-V}$/$\nhsym$=1.7$\times$10$^{-22}$ mag cm$^{2}$ \citep{Boh78}. 
In the fitting procedure, we allowed this component to  
 vary in the range of 0.5--5$\times$10$^{21}$
cm$^{-2}$. The third component is a neutral absorber  partially covering the
central source  ({\it zpcfabs} model in {\textsc{xspec}}),  associated with  material on the
torus distance scale. The full parametrization adopted is the following one: {\it
phabs$\times$(zphabs$\times$zpcfabs$\times$zpowerlw)}. 
We allowed to vary the intrinsic power-law normalization and both 
the intrinsic column density and the covering factor
of the partial covering absorber.
The residuals between
the data and the best-fitting model for the two states are shown in Fig.~\ref{fig4}. 

\begin{figure}
\centering
\hskip -0.7cm\includegraphics[width=6cm, height=9cm, angle=-90]{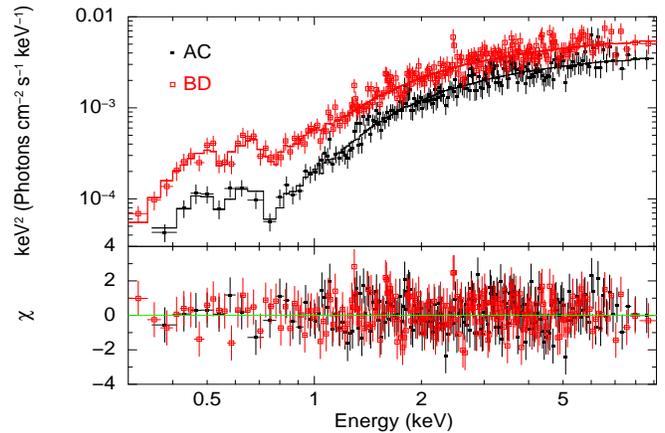}
\caption{Upper panel: The model including an ionized absorber partially covering the
source ({\it phabs}$\times$({\it zphabs}$\times${\it zxipcf}$\times${\it zpowerlw}))
is overplotted on the low- (AC) and high-state (BD)
unfolded spectra of Mrk 915.
Lower panel: Relevant residuals, plotted in terms of sigmas with error bars of size one.}
\label{fig5}
\end{figure}

Although this model can account for most of the data, it leaves 
 evident and systematical residuals in the
softer part of the  spectra: by considering the data in the
0.3--1 keV range we found $\chi$$^2/$ d.o.f.=55.42/37 (corresponding to
a null hypothesis probability of $\sim$3 per cent). Thus, this model can not be considered a good 
representation of the global spectral
properties of the source.  In particular, the shape of the residuals in  Fig.~\ref{fig4}
suggests the presence of an absorption trough at $\sim$0.78 keV (observed
frame); at the redshift of the source, this feature could be associated with 
{Fe\,\textsc{xvii}-\,\textsc{xviii}}, typically 
observed in nearby bright AGN with warm absorbers \citep{Cre03,Por04}. Prompted by these considerations,  we replaced the neutral partial
covering component with an ionized one, modelled in \textsc{XSPEC} with \textit{zxipcf}
\citep{Ree08}. In this case, the full parametrization  is the following:
{\it phabs$\times$(zphabs$\times$zxipcf$\times$zpowerlw)}. 
As a first step, in the fitting
procedure, we left the intrinsic $\nhsym$, the ionization parameter 
$\xi${\footnote{The ionization parameter
is defined as $\xi$[erg cm s$^{-1}$]= $L_{\rm{ion}}$/\textit{nR}$^{2}$ \citep{Tar69}, where
$L_{\rm{ion}}$ is the ionization luminosity obtained by integrating the X-ray intrinsic luminosity 
between 0.013 and 13 keV,
\textit{n} is the average absorber number density [part/cm$^{-3}$] of the illuminated
slab, and \textit{R} is the distance of the absorber from the central source.} and the covering factor
(CF) of the ionized medium free to vary,
along with the intrinsic power-law normalizations ($A_{\rm{p}}$).}
The best-fitting values obtained
in the presence of an ionized absorber are reported in  Table~\ref{model}, where  the 
uncertainties are given at the 90 per cent confidence level for one parameter of
interest \citep{Avn76}. 
The unfolded spectra
and the ratio between data and best-fitting models are shown in Fig.~\ref{fig5}; it is
evident that  this second model provides a better characterization of the spectra of
Mrk 915, also in the softer energy range
($\chi$$^2/$d.o.f.=22.42/35 in the 0.3--1 keV range, corresponding 
to a null hypothesis probability of $\sim$95 per cent). Also the 
residuals at $E\sim$0.78 keV are now well reproduced. We note that, 
with the present data, we cannot investigate the presence of a
possible outflow velocity. Thus, the {\it redshift} 
of the ionized absorber was fixed to
the same {\it z} of
the source.

\begin{figure}
\includegraphics[width=8cm]{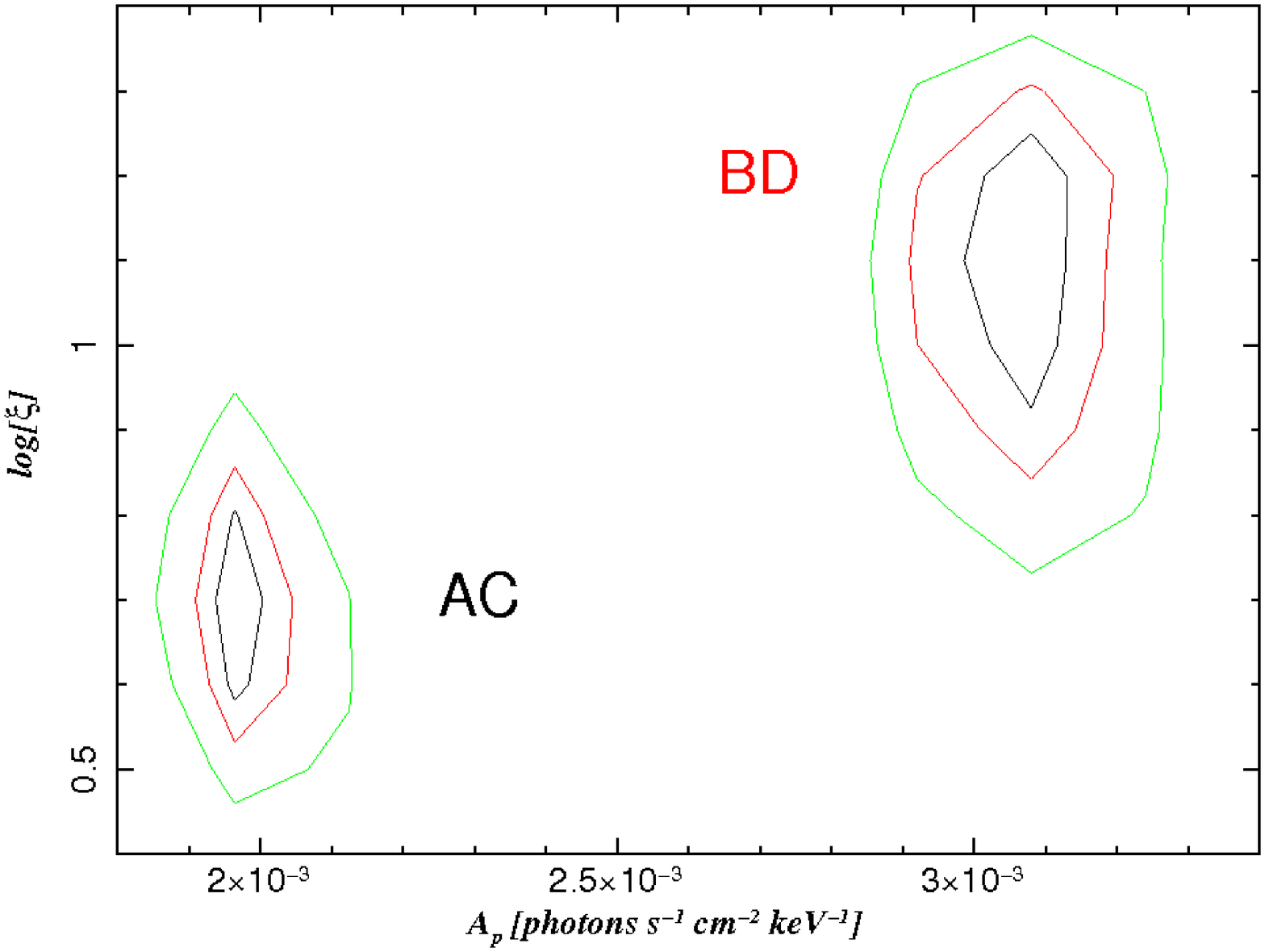}
\includegraphics[width=8cm]{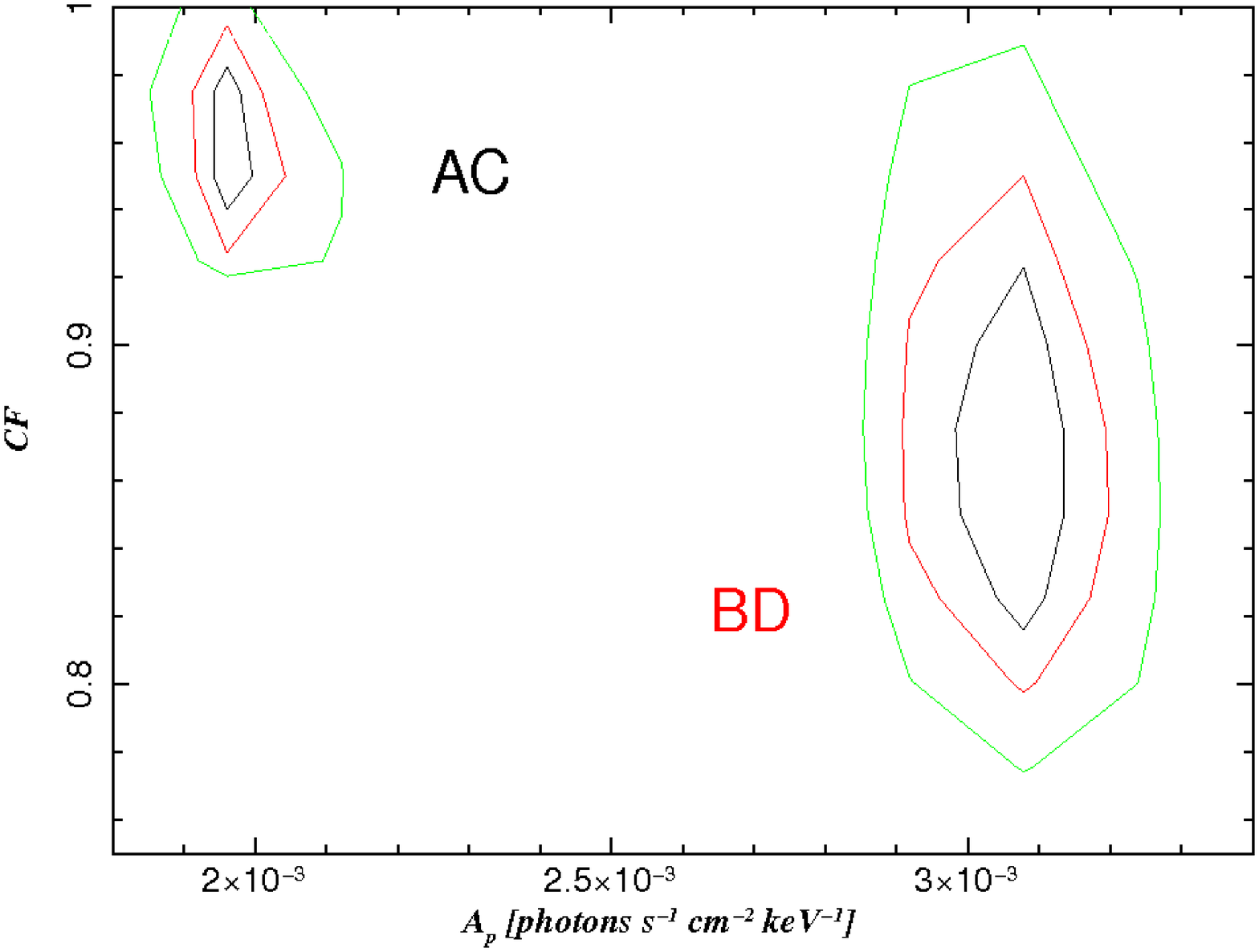}
\includegraphics[width=8.3cm]{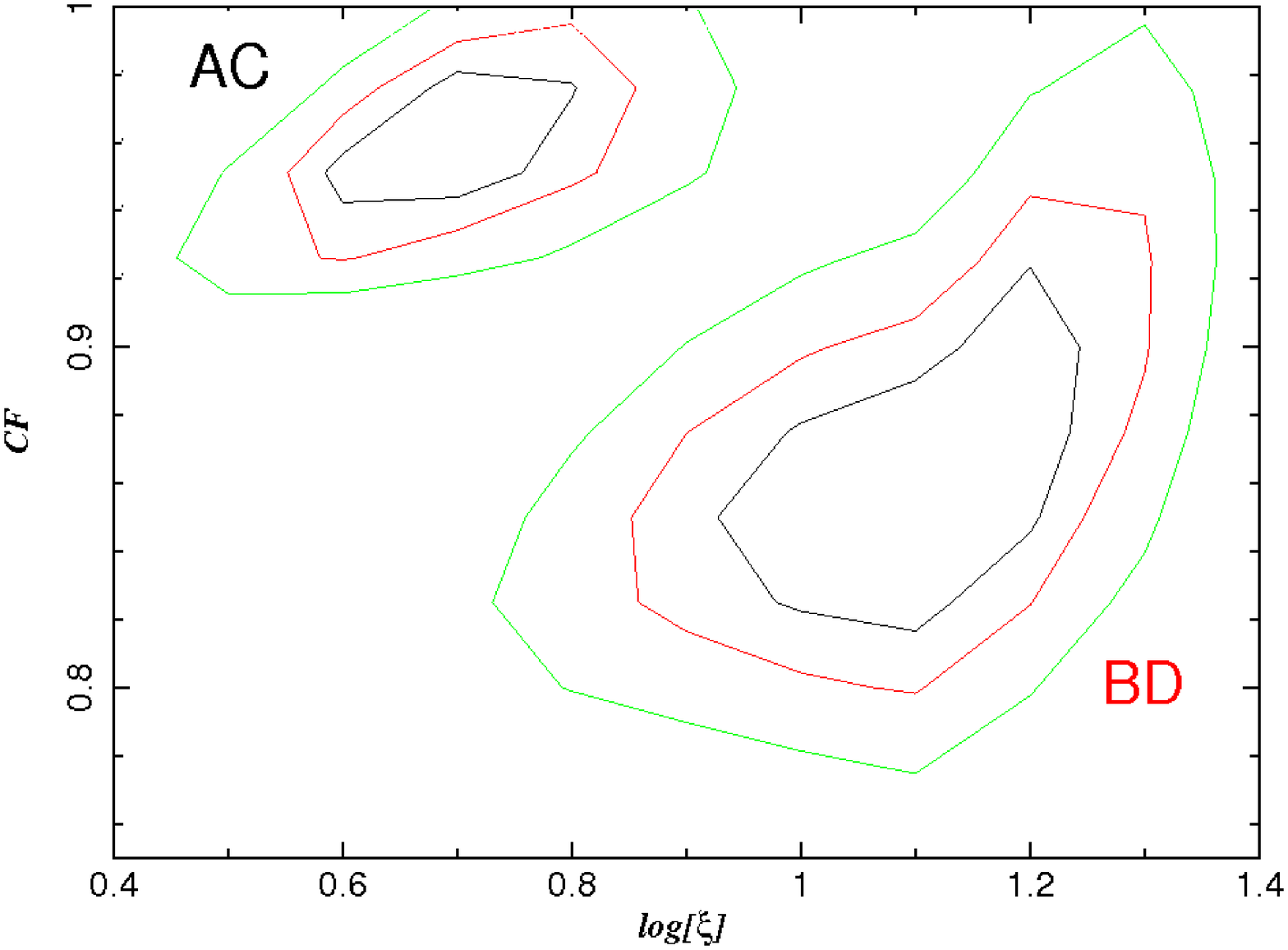}
\caption{68 per cent, 90 per cent and 99 per cent confidence contours for the AC state and
BC states  for different parameters of interest obtained by using the model reported in Table~\ref{model}.
Top panel: ionization state versus the power--law  normalization ($A$$_{\rm{p}}$).
Middle panel:  Covering factor versus the power--law  normalization ($A$$_{\rm{p}}$).
Bottom panel: Covering factor versus the  ionization state.}
\label{fig6}
\end{figure}

The confidence contour plots of the joint errors for
different parameters of interest (log$(\xi)$, CF
and  $A_{\rm{p}}$), obtained with the same choices of fixed parameters as quoted in Table~\ref{model},  
are shown in Fig.~\ref{fig6}. 
The emerging scenario is that, in both states, the nucleus is obscured by 
a variable ionized medium with an  $\nhsym$ of the order of 
2$\times$10$^{22}$ cm$^{-2}$ and a significant increase in the 
power-law  normalization by a  factor of $\sim$1.5 from 
the low- (AC) to the high-state (BD)  is also required (see Table~\ref{model}).
We note that, this increase is required also in the case of neutral partial covering absorber.
In combination with an intrinsic fading/increasing of the power--law 
normalization, Fig.~\ref{fig6} suggests also  a marginally evidence for a change  of
the ionization state and of the CF, albeit with low significance.
In
particular, when the source gets brighter  the ionization state tentatively
increases (90 per cent confidence level, see Fig.~\ref{fig6}, top panel)  and the CF
slightly decreases (68 per cent confidence level, see Fig.~\ref{fig6},  middle panel).

In the bottom panel of Fig.~\ref{fig6}, we report the 
confidence contour plot of the joint errors of the CF versus the ionization
parameter.  This plot indicates that our spectra could be equally reproduced
 by assuming that only one of these two parameters is changing.
Thus, we re-fitted the data twice,  forcing the two spectral states to have: (1)
the same ionization parameter, allowing the CF to be free to vary, and, (2)  the
same CF, allowing  the ionization parameter to be free to vary. As expected, in
both the cases we were able to reproduce our spectra with a statistical quality 
similar to that reported in Table~\ref{model} and showed in Fig.~\ref{fig5}. 
In both cases, the intrinsic column density for the ionized medium 
($\nhsym$$\sim$2$\times$10$^{22}$ cm$^{-2}$) and the
change in the power-law normalization (by a factor of 1.5) are consistent with the values previously
obtained. 

In the case in which the two ionization parameters were tied together (log($\xi$)=0.63$\pm{0.15}$),
the spectral shape  variation could be completely ascribed to a change in the CF from
$\sim$0.97 in the low-spectral state to $\sim$0.85 in the high-spectral state ($\chi$$^2/$d.o.f.=365.33/404). In the case with the two covering factors tied 
together (CF=0.95$\pm{0.03}$), the spectral shape  variation could be completely ascribed to a change in
the log($\xi$) from  $\sim$0.77 in the low-spectral state to $\sim$1.35 in the high-spectral state
($\chi$$^2/$d.o.f.=363.5/404).

Finally, we tested the picture discussed above fitting our final model,
with both  hypotheses of constant $\xi$ or constant CF, to the single spectra
showed in Fig.~\ref{fig3} (i.e. A, B, C, and D spectra spaced by about 5$\times$10$^{5}$ s, see Fig.~\ref{fig1}). We found  results
consistent with those previously described.  In spite of the lower statistics of the data with respect
to the combined spectra, the variations in the $\xi$ and CF parameters still remains
statistically significant (68--90 per cent confidence level) as the change in the power-law 
normalizations (more than 99 per cent confidence level).

\section{Discussion}

The spectral analysis performed on the XRT monitoring data of Mrk 915 
unveiled the presence of an ionized absorber with $\nhsym$ of
the order of 2$\times$10$^{22}$ cm$^{-2}$ and allowed us to detect a variability of the
intrinsic power of the central source of a factor of $\sim$1.5 on a time-scale of a few
days.  The possibility that the source could intrinsically decrease/increase its power was
already suggested from observations in the optical domain (see Section 2 for relevant references).

As for the spectral variation  observed in the X-ray domain, we found that it could be due to  a change in the ionization state of  the
circumnuclear medium and/or to a change of the covering of absorbing medium along the line of sight.
Higher  quality data are needed to understand if only one or both
these parameters actually vary.
If the variation of the
covering factor will be confirmed, this implies a clumpy structure of 
the absorber.
Since our X-ray data can not firmly confirm or discard this hypothesis, in the following 
we discuss some constraints on the location of the absorber
by assuming both a homogeneous and a clumpy medium. 

{\it Homogeneous medium.} Under the hypothesis of  homogeneous medium,  we
can place an upper  limit on the location of this absorber by using the
relation between the intrinsic continuum luminosity, the ionization and the
density of the medium: $\xi$ = $L_{\rm{ion}}$/{\it nR$^{2}$}. Assuming that the
thickness of the absorber is less than the distance from the central SMBH
($\Delta$R/R$<$ 1),   this relation provides an upper limit on its distance:
{\it R}$<$$L_{\rm{ion}}$/$\nhsym$$\xi$. From the X-ray spectral analysis we
derived: $\nhsym$$\sim$2$\times$10$^{22}$ cm$^{-2}$ and
$L_{\rm{ion}}$=(2.4--3.7)$\times$10$^{43}$erg s$^{-1}$ (low-high state,
respectively; $L_{\rm{ion}}$ is the intrinsic X-ray luminosity, corrected for absorption, and integrated between
0.013 and 13 keV, see also footnote 7). 
By considering the minimum value of $\xi$ obtained under the
hypothesis of no covering factor variation, we obtain
$R$$<$(2--3)$\times$10$^{20}$ cm,  i.e. lower than (0.8--3)$\times$10$^7$ $R_{\rm{g}}$
({\it R$_{\rm{g}}$}={\it GM$_{\rm{BH}}$/c$^2$} is the gravitational radius, where
$M_{\rm{BH}}$ is the SMBH mass equal to 0.6-1.8$\times$10$^8$ M$_{\odot}$;
\citealt{Ben06}). This estimate corresponds to a location inside the
outer edge of the torus.

{\it Clumpy medium.} Under the hypothesis of a clumpy medium and  assuming 
that the absorber is composed by spherical gas 
clouds moving around the central source with Keplerian velocity (e.g. 
\citealt{Ris02}, \citealt{Elv04}, \citealt{Ris05}), the variability time-scale between the two maxima (B and D) 
detected in
Fig.~\ref{fig1} and Fig.~\ref{fig2}, {\it $<$t$>$}=10$^{6}$ s, at the first order, is equal 
to the crossing time for a single  cloud
through the source.  

The distance {\it R$_{\rm{c}}$} of the absorber from the central source is  given by: 
{\it R$_{\rm{c}}$}={\it GM$_{\rm{BH}}$ t$_{\rm{c}}$$^{2}$ / ({\it D$_{\rm{s}}$}+{\it D$_{\rm{c}}$})$^{2}$}
 where {\it t$_{\rm{c}}$} is the variability time and {\it D$_{\rm{c}}$} and {\it D$_{\rm{s}}$}
are the cloud and central source linear size, respectively.
Under the hypothesis that the linear size of a single 
cloud is of the same order of that of the emitting central
source, and taking into account that {\it D$_{\rm{c}}$}=$\nhsym$$_{\rm{, c}}$/{\it n$_{\rm{c}}$},
we can re-write the distance {\it R$_{\rm{c}}$} in the following way:
{\it R$_{\rm{c}}$}= 0.25~{\it GM$_{\rm{BH}}$ t$_{\rm{c}}$$^{2}$n$_{\rm{c}}$$^{2}$ $\nhsym$$_{\rm{, c}}$$^{-2}$}, i.e.,
for $\nhsym$$_{\rm{, c}}$=2$\times$10$^{22}$ cm$^{-2}$:
\begin{equation}
\hskip 3cm{\it R_{\rm{c}}}\sim5-15~{\it n_{\rm{c}}^{2}}~\mbox{cm}
\end{equation}
The range of values reported in equation (1) is due to the uncertainty in the 
$M_{\rm{BH}}$ (0.6--1.8$\times$10$^8$ M$_{\odot}$, see Section~2).

A conceptually different estimate of {\it R$_{\rm{c}}$} can be derived
from the relation quoted above 
between the intrinsic continuum luminosity, the ionization parameter 
and the
density of the absorber ($\xi$ = $L_{\rm{ion}}$/{\it n$_{\rm{c}}$R$_{\rm{c}}^{2}$}).
By considering an average value of the
$\xi$ parameter of 10 erg cm s$^{-1}$ and of the intrinsic ionization
luminosity of $L_{\rm{ion}}$$\sim$3$\times$10$^{43}$ erg s$^{-1}$, we can derive:
\begin{equation}
\hskip 2.5cm{\it R_{\rm{c}}}=(3\times10^{42} n_{\rm{c}}^{-1})^{1/2} \mbox{cm.}
\end{equation}
By combining the results reported in (1) and (2), we find that, under the
hypothesis of absorbing clouds with linear dimension similar to that 
of the central source, the clumpy medium 
should be located at a distance 
{\it R$_{\rm{c}}$}=(1.4--1.7)$\times$10$^{17}$ cm (i.e. $\sim$10$^4${\it R$_{\rm{g}}$}), 
with an average cloud density of
{\it n$_{\rm{c}}$}$\sim$(1--1.7)$\times$10$^{8}$ cm$^{-3}$ and an average linear dimension of
{\it D$_{\rm{c}}$}$\sim${\it D$_{\rm{s}}$}$\sim$(1--1.2)$\times$10$^{14}$ cm (i.e. $\sim$7{\it R$_{\rm{g}}$}).

We repeated the estimates discussed above by
taking into account that the linear size of the single cloud could indeed be
smaller (e.g. {\it D$_{\rm{c}}$}$\sim$0.5~{\it D$_{\rm{s}}$}) or greater ({\it D$_{\rm{c}}$}$>>${\it D$_{\rm{s}}$}) 
with respect to that of the central source.
We found that {\it R$_{\rm{c}}$} could vary between $\sim$10$^{17}$ cm and 
2.2$\times$10$^{17}$ cm with a density {\it n$_{\rm{c}}$} between
2.2$\times$10$^{8}$ cm$^{-3}$ and 6$\times$10$^{7}$ cm$^{-3}$, respectively.

By considering the   5100$\AA$ luminosity estimated by 
\citet{Ben06} for this source  ($\lambda L_{\lambda}\sim1.68\times10^{44}$ erg
s$^{-1}$)  and the relation found by \citet{Ben09} between the BLR size and
the 5100$\AA$ luminosity, the BLR in Mrk 915 is expected to be located at a distance
from the centre SMBH of the order of 10$^{16}$-- 10$^{17}$ cm. 
If confirmed, the short time-scale  variation of the CF would imply 
that the ionized absorber clouds should be located near,
or just outside, to the bulk of the  BLR with a density between 
 6$\times$10$^{7}$  and 2.2$\times$10$^{8}$ cm$^{-3}$.
This range of densities would indeed be consistent with the expected 
lower limit of the electron density for the
BLR ({\it n$_{\rm{c}}$}(BLR)$>$10$^{8}$ cm$^{-3}$; \citealt{Ost89}).
We note that this scenario is very similar to that proposed by \citet{Goo95},
see Section 2,
to explain the optical variability of this source.

\section{Summary and Conclusions}
From the optical point of view, Mrk~915 is a spiral galaxy hosting
an intermediate (type 1.5--1.9) AGN with evident
dust lane structures crossing the central source.
Significant variability (about a factor of 2.5--3) in the intensity
of both the broad component of the H$\alpha$ emission line and the
underlying continuum
has been observed and reported in the literature on time-scale of few years.
The analyses carried out by different authors suggest that this change
could be due to an increasing of the reddening along the line of sight, to a fading of the central source,
or to a combination of them.
Thanks to the unique capabilities of XRT on-board the {\it Swift} satellite, 
it was possible to investigate the daily time-scale variation of the X-ray
absorber.

As discussed in Section 1,  Mrk 915 was selected as a good candidate for
a variable absorber on the basis of the X-ray colour analysis  on archival XRT
data. The analysis of these observations \citep{Bal15}, 
which caught the source in  two different  states, suggested that this
could be associated with a variable absorber.

With the aim at obtaining higher quality data on shorter time-scale, we
performed a XRT monitoring covering $\sim$3 weeks and sampling days time-scale. 
During these new observations the source was caught again in two 
main different count rate and spectral
states. We investigated the nature of the variable absorber testing two partial
covering models: in the first one
the absorber is neutral, while in the second one it is ionized. In both
cases, we added a neutral absorber ($\nhsym$$\sim$10$^{21}$ cm$^{-2}$) 
associated with the
dust lanes structures seen crossing the central source.
We find that, while the first model (neutral absorber) is not able to reproduce the
data below 1 keV, the second one  (ionized absorber) 
provides a good characterization of the data in the full energy range considered here. 
Furthermore, we detect a significant
variation  of the intrinsic power of the nuclear source.
The quality of the present data prevents us from firmly establishing if
the observed spectral variations were due to a change in the ionization state of 
the circumnuclear medium and/or to its covering factor.
Higher  quality data are needed to establish if only one or both of these parameters change
in this source. 
By combining the information provided by the X-ray
light curve and by the spectral analysis,
we derived some constraints on the absorber location under the hypothesis of
a homogeneous and a clumpy medium.
In both cases, the absorber should be located inside the outer edge of an extended torus and, in particular, 
under the clumpy hypothesis, it should be located close, or just outside, to the BLR zone.

\section*{Acknowledgments}

This work is based on observations obtained with the {\it Swift} satellite and it made use of data supplied 
by the UK {\it Swift} Science Data Centre at the University of Leicester. We thank Neil Gehrels, Boris Sbarufatti, Gianpiero Tagliaferri 
and the {\it Swift} Mission Operation Center to make every effort to get our ToO request scheduled.  
This research has made use both of the Palermo BAT Catalogue and database operated at INAF - IASF Palermo 
and of the 70 month {\it Swift}-BAT Catalogue. 
We, moreover,  thank Miguel Perez-Torres for the analysis
of archival VLA data. 
Part of this work was supported by the European Commission Seventh Framework Programme (FP7/2007-2013) 
under grant agreement no. 267251 Astronomy Fellowships in Italy (AstroFIt). The authors acknowledge
financial support from the Italian Ministry of Education,
Universities and Research (PRIN2010-2011, grant no.
2010NHBSBE). Support from the Italian Space Agency is
acknowledged (contract ASI INAF I/037/12/0). 
Finally, we you would like to thank the anonymous referee for the useful and constructive comments which improved the quality of the paper.

\label{lastpage}

\begin{thebibliography}{99}
\bibitem[\protect\citeauthoryear{Antonucci}{1993}]{Ant93} Antonucci R., 1993, ARA\&A, 31, 473
\bibitem[\protect\citeauthoryear{Avni}{1976}]{Avn76} Avni Y., 1976, ApJ, 210, 642 
\bibitem[\protect\citeauthoryear{Ballo et al.}{2015}]{Bal15} Ballo L. et al., 2015, ed. by P. Caraveo, P. D'Avanzo, N. Gehrels and G. Tagliaferri;
Proc. of ``Swift: 10 years of discovery", 2-5 December 2014, Rome,
Italy; Proceedings of Science (PoS, Trieste, Italy), vol. SWIFT 10, PoS(SWIFT10)122
(arXiv:1505.02593)
\bibitem[\protect\citeauthoryear{Baumgartner et al.}{2013}]{Bau13} Baumgartner W. H., Tueller J., Markwardt C. B.,
Skinner G. K., Barthelmy S., Mushotzky R. F., Evans P. A., Gehrels N., 2013, ApJS, 207, 19
\bibitem[\protect\citeauthoryear{Bennert et al.}{2006}]{Ben06} Bennert N., Jungwiert B., Komossa S., Haas M., Chini R., 2006, A\&A, 459, 55 
\bibitem[\protect\citeauthoryear{Bentz}{2009}]{Ben09} Bentz M.~C., Peterson B.~M., Netzer H., Pogge R.~W., Vestergaard M., 2009, ApJ, 697, 160 
\bibitem[\protect\citeauthoryear{Bianchi, Maiolino, \& Risaliti}{2012}]{Bia12} Bianchi S., Maiolino R., Risaliti G., 2012, Adv. Astron., 2012, 782030 
\bibitem[\protect\citeauthoryear{Bohlin, Savage, \& Drake}{1978}]{Boh78} Bohlin R.~C., Savage B.~D., Drake J.~F., 1978, ApJ, 224, 132 
\bibitem[\protect\citeauthoryear{Braito et al.}{2014}]{Bra14}  Braito V., Reeves J.~N., Gofford J., Nardini E., Porquet D., Risaliti G., 
2014, ApJ, 795, 87 
\bibitem[\protect\citeauthoryear{Burrows et al.}{2005}]{Bur05} Burrows D.~N., et al., 2005, Space Sci. Rev., 120, 165 
\bibitem[\protect\citeauthoryear{Connolly, McHardy, \& Dwelly}{2014}]{Con14} Connolly S.~D., McHardy I.~M., Dwelly T., 2014, MNRAS, 440, 3503 
\bibitem[\protect\citeauthoryear{Crenshaw, Kraemer, \& George}{2003}]{Cre03} Crenshaw D.~M., Kraemer S.~B., George I.~M., 2003, ARA\&A, 41, 117 
\bibitem[\protect\citeauthoryear{Cusumano et al.}{2010}]{Cus10} Cusumano G. et al.,  2010, A\&A, 524, A64 
\bibitem[\protect\citeauthoryear{Elvis et al.}{2004}]{Elv04} 	Elvis M. Risaliti G., Nicastro F., Miller J. M., 
Fiore F., Puccetti S., 2004, ApJ 615, L25
\bibitem[\protect\citeauthoryear{Evans et al.}{2007}]{Eva07} Evans P.~A. et al., 2007, A\&A, 469, 379 
\bibitem[\protect\citeauthoryear{Evans et al.}{2009}]{Eva09} Evans P.~A. et al., 2009, MNRAS, 397, 1177 
\bibitem[\protect\citeauthoryear{Goodrich}{1995}]{Goo95} Goodrich R.~W., 1995, ApJ, 440, 141
\bibitem[\protect\citeauthoryear{Kalberla et al.}{2005}]{Kal05} Kalberla P.~M.~W., Burton W.~B., Hartmann D., Arnal E.~M., Bajaja E., Morras R., P{\"o}ppel W.~G.~L., 2005, A\&A, 440, 775 
\bibitem[\protect\citeauthoryear{Keel}{1996}]{Kee96} Keel W.~C., 1996, ApJS, 106, 27 
\bibitem[\protect\citeauthoryear{Malkan, Gorjian, \& Tam}{1998}]{Mal98} Malkan M.~A., Gorjian V., Tam R., 1998, ApJS, 117, 25 
\bibitem[\protect\citeauthoryear{Moretti et al.}{2005}]{Mor05} Moretti, A.  et al. 2005, in Siegmund O. H. W., ed., Proc. SPIE Conf. Ser. Vol. 5898, UV, X-ray, and Gamma-Ray
Space Instrumentation for Astronomy XIV. SPIE,  Bellingham, p. 360
\bibitem[\protect\citeauthoryear{Mu{\~n}oz Mar{\'{\i}}n et al.}{2007}]{Mun07} Mu{\~n}oz Mar{\'{\i}}n V.~M., Gonz{\'a}lez 
Delgado R.~M., Schmitt H.~R., Cid Fernandes R., P{\'e}rez E., Storchi-Bergmann T., Heckman T., Leitherer C., 2007, AJ, 134, 648 
\bibitem[\protect\citeauthoryear{Osterbrock}{1989}]{Ost89} Osterbrock D.~E., 1989, Sky Telesc., 78, 491 
\bibitem[\protect\citeauthoryear{Porquet et al.}{2004}]{Por04} Porquet D., Reeves J.~N., O'Brien P., Brinkmann W., 2004, A\&A, 422, 85 
\bibitem[\protect\citeauthoryear{Reeves et al.}{2008}]{Ree08} Reeves J., Done C., Pounds K., Terashima Y., Hayashida K., Anabuki N., 
Uchino M., Turner M., 2008, MNRAS, 385, L108 
\bibitem[\protect\citeauthoryear{Risaliti et al.}{2002}]{Ris02} Risaliti G., Elvis M., Nicastro F., 2002, ApJ, 571, 234 
\bibitem[\protect\citeauthoryear{Risaliti et al.}{2005}]{Ris05} Risaliti G., Elvis M., Fabbiano G., Baldi 
A., Zezas A., 2005, ApJ, 623, L93 
\bibitem[\protect\citeauthoryear{Risaliti et al.}{2007}]{Ris07} Risaliti G., Elvis M., Fabbiano G., Baldi 
A., Zezas A., Salvati M., 2007, ApJ, 659, L111 
\bibitem[\protect\citeauthoryear{Risaliti et al.}{2009}]{Ris09} Risaliti, G., Salvati, M., Elvis, M., Fabbiano, G., 
Baldi, A., Bianchi, S., Braito, V., Guainazzi, M., Matt, G., Miniutti, G., Reeves, J., Soria, R., Zezas, A., 
2009, MNRAS 393, L1
\bibitem[\protect\citeauthoryear{Tarter, Tucker, \& Salpeter}{1969}]{Tar69} Tarter C.~B., Tucker W.~H., Salpeter E.~E., 1969, ApJ, 156, 943 
\bibitem[\protect\citeauthoryear{Trippe et al.}{2010}]{Tri10} 
Trippe M.~L., Crenshaw D.~M., Deo R.~P., Dietrich M., Kraemer S.~B., Rafter 
S.~E., Turner T.~J., 2010, ApJ, 725, 1749 
\bibitem[\protect\citeauthoryear{Whittle}{1992}]{Whi92} Whittle M., 1992, ApJS, 79, 49 
\end{thebibliography}
\end{document}